\documentclass[fleqn,twoside]{article}
\usepackage{espcrc2}
\usepackage{epsfig}
\usepackage{latexsym}
\usepackage{amssymb}
\usepackage{amsfonts}

\readRCS
$Id: espcrc2.tex,v 1.2 2004/02/24 11:22:11 spepping Exp $
\usepackage{graphicx}
\usepackage[figuresright]{rotating}

\newcommand{\AmS}{{\protect\the\textfont2
  A\kern-.1667em\lower.5ex\hbox{M}\kern-.125emS}}

\title{Evaluating the three-loop static quark potential}

\author{Alexander V.~Smirnov\address{Scientific Research Computing Center of
Moscow State University, Moscow 119992, Russia}$^c$,
Vladimir A.~Smirnov\address{Nuclear Physics Institute of
  Moscow State University,
  Moscow 119992, Russia}$^c$\thanks{Talk given at the International
  Workshop `Loops and Legs in
  Quantum Field Theory' (April 20--25, 2008, Sondershausen, Germany).}
and Matthias Steinhauser\address{Institut f{\"u}r Theoretische Teilchenphysik,
  Universit{\"a}t Karlsruhe (TH),
  76128 Karlsruhe, Germany}}

\newcommand{\ep}{\varepsilon}

\newcommand{\be}{\begin{equation}}
\newcommand{\ee}{\end{equation}}
\newcommand{\bea}{\begin{eqnarray}}
\newcommand{\eea}{\end{eqnarray}}

\newcommand{\Gm}{\Gamma}

\newcommand{\dd}{\mbox{d}}

\newcommand{\nn}{\nonumber}

\begin{document}

\begin{abstract}
This is a status report of the evaluation of the three-loop corrections to the
static QCD potential of a heavy quark and an antiquark.
The families of Feynman integrals that appear in the evaluation are
described. To reduce any integral of the families to master integrals
we solve integration-by-parts relations by the algorithm
called {\tt FIRE}. To evaluate the corresponding master integrals we apply
the Mellin--Barnes technique. First results are
presented: the coefficients of $n_l^3$ and $n_l^2$,
where $n_l$ is the number of
light quarks.
\vspace{1pc}
\end{abstract}

\maketitle

\sloppy


\section{Introduction}

The QCD potential between a static quark and its antiquark can be cast in the
form
\begin{eqnarray}
  V(|{\vec q}\,|) &=& -{4\pi C_F\alpha_s\over{\vec q\,}^2}
  \Bigg[1+{\alpha_s\over 4\pi}a_1
  +\left({\alpha_s\over 4\pi}\right)^2a_2
  \nn \\ &&  \hspace*{-12mm}
  +\left({\alpha_s\over 4\pi}\right)^3
  \left(a_3+ 8\pi^2 C_A^3\ln{\mu^2\over{\vec q\,}^2}\right)
  +\cdots\Bigg]\,,
  \label{singlet}
\end{eqnarray}
where the renormalization scale of $\alpha_s$ is set to $\vec{q\,}^2$.
The one-loop contribution $a_1$ is known since almost 30
years~\cite{Fischler:1977yf,Billoire:1979ih}
and also the two-loop term has already been computed end of the nineties
\cite{Peter:1996ig,Peter:1997me,Schroder:1998vy}. Furthermore
logarithmic contributions are known at three- and four-loop
level~\cite{Pineda:2000gza,Brambilla:2006wp}.
Explicit results are nicely summarized in
the recent review~\cite{Vairo:2007id}.
The non-logarithmic third-order term, $a_3$, is still unknown.


\section{Reduction to master integrals}

\begin{figure}[t]
  \centering
  \leavevmode
  \epsfxsize=.4\textwidth
  \epsffile[130 280 530 550]{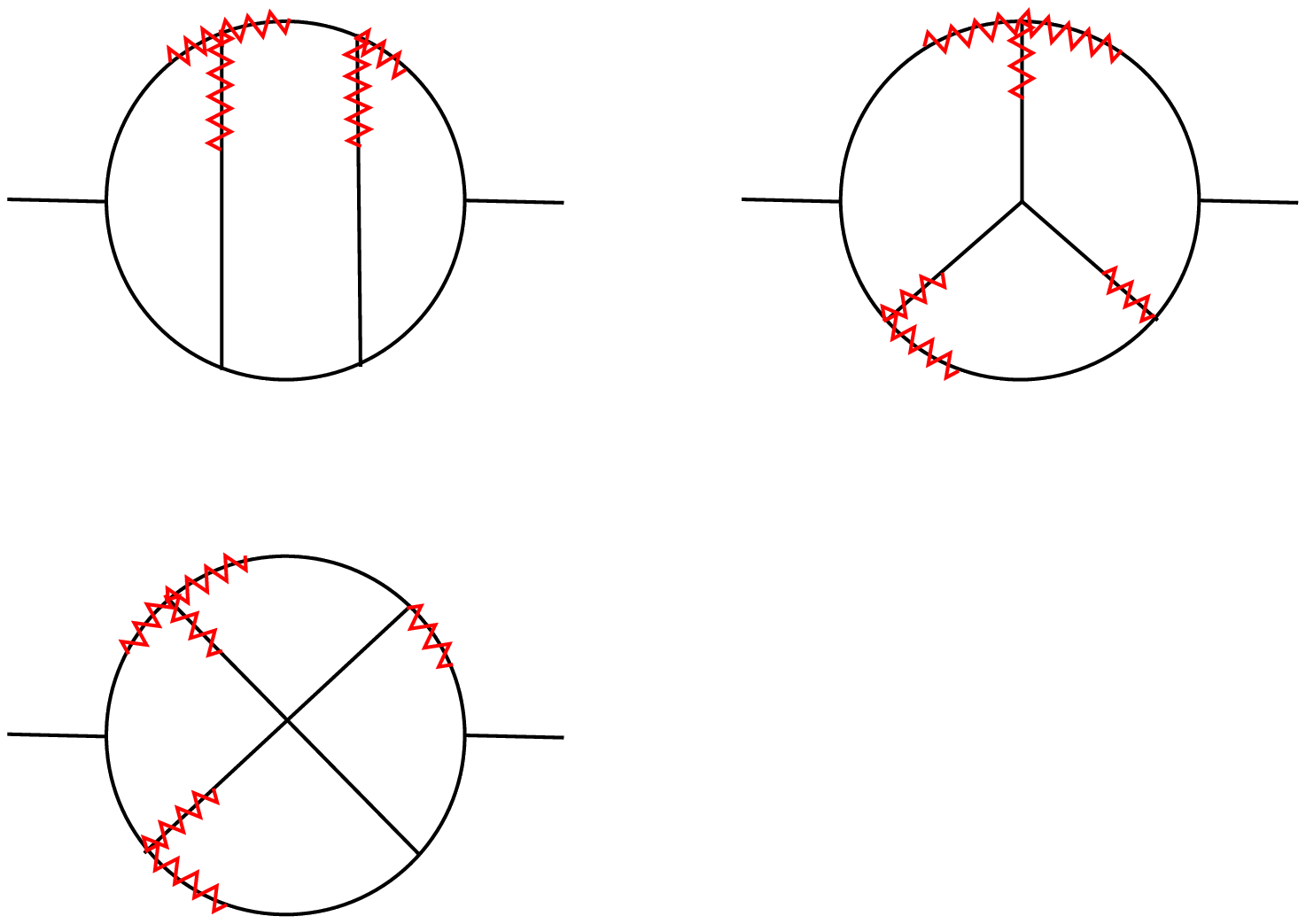}
  \caption{\label{BasicGraphs}
    Families of Feynman integrals needed for the calculation of $a_3$. The
    solid lines correspond to relativistic massless propagators and the zigzag
    lines represents static propagators.}
\end{figure}

Any Feynman integral that contributes to $a_3$ can be mapped to
one of the three graphs shown in Fig.~\ref{BasicGraphs}
where solid lines stand
for usual massless propagators of the form
$1/(-p^2-i0)^{a_i}$
and wavy lines stand for linear propagators $1/(-v\cdot p-s_i i0)^{a_i}$,
with $v\cdot p=p_0$, $s_i=\pm 1$ and integer indices $a_i$.
In case the latter type of propagators is absent the integrals reduce to
usual massless two-point functions which can be treated with the help of
{\tt MINCER}~\cite{Larin:1991fz}. Note, however, that the presence of the
static lines significantly increases the complexity of the problem.

In general the integrals involve up to fifteen propagators (including an
irreducible numerator). In order to
simplify the reduction problem we apply in a first step
partial fraction identities to arrive at various subfamilies of integrals with
at most three linear propagators.
Thus any resulting integral is labeled by twelve indices one of which
corresponds to an irreducible numerator and three indices correspond to linear
propagators.

\begin{figure}[b]
  \centering
  \leavevmode
  \epsfxsize=.4\textwidth
  \epsffile[120 320 560 490]{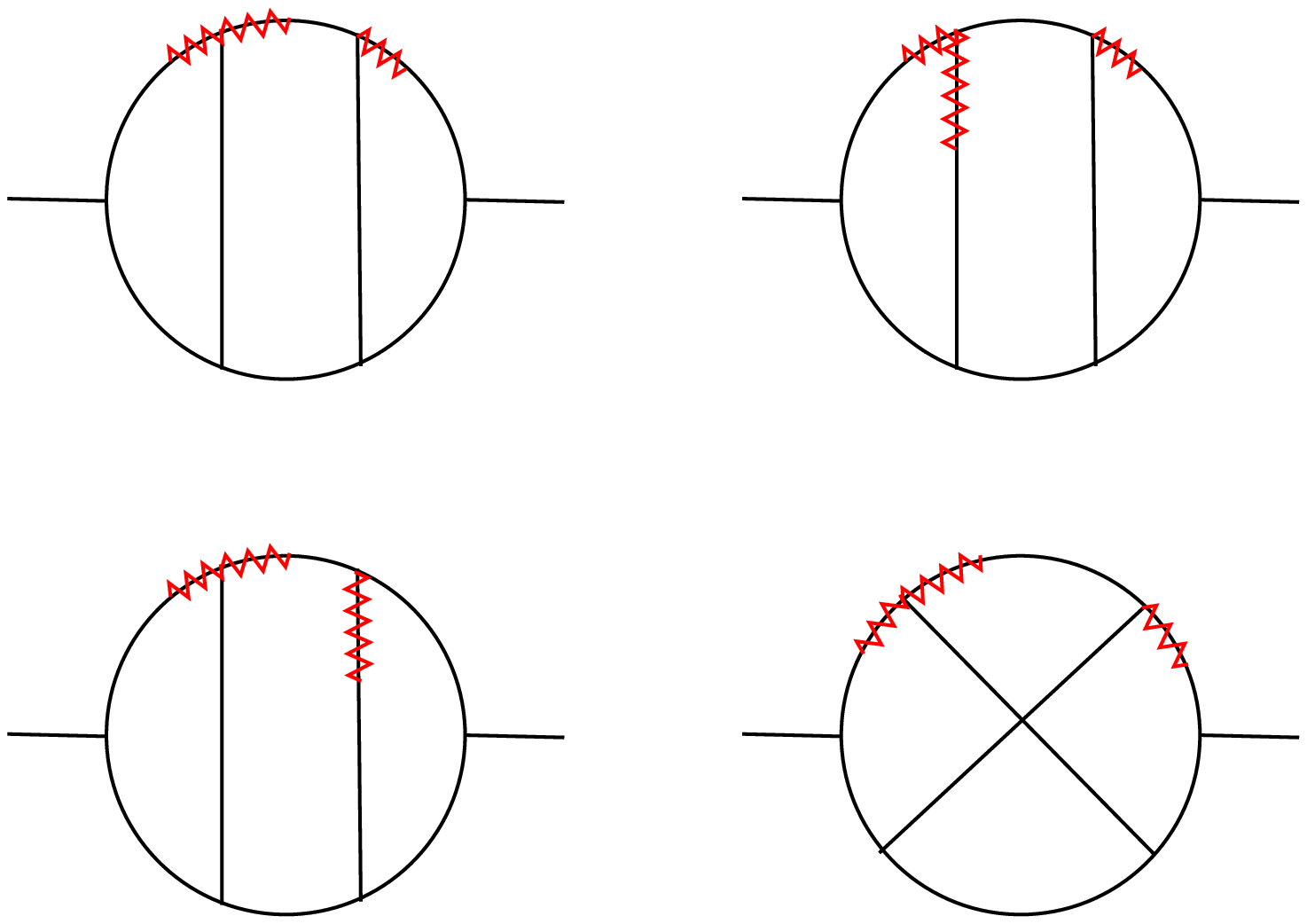}
  \caption{\label{LaNonPl}
    Ladder and non-planar diagrams contributing to the $n_l$ part of
    $a_3$.}
\end{figure}

Let us in the following describe the Feynman integrals that are generated by
the $n_l$ contribution. Altogether we have to consider about 70\,000 integrals
(allowing for a general QCD gauge parameter $\xi$).
As far as the ``ladder'' and ``non-planar'' diagrams are concerned we
have to deal with the type of integrals shown in Fig.~\ref{LaNonPl} where the
linear propagators appear in the following form: If the loop momenta, $k$,
$l$ and $r$, in Fig.~\ref{LaNonPl} are chosen as the momenta of the three
upper lines, then the first diagram appears in two ways:
either with the product
\[(-v\cdot k-i0)^{-a_9}(-v\cdot l-i0)^{-a_{10}} (-v\cdot r-i0)^{-a_{11}},\]
or with the product
\[(-v\cdot k+i0)^{-a_9}(-v\cdot l-i0)^{-a_{10}} (-v\cdot r-i0)^{-a_{11}}.\]
The second diagram appears with similar propagators where
the momenta $\{k,l,r\}$ (in the first variant with $-i0$ in three places) are
replaced by $\{k,k-l,r\}$.
The third diagram in Fig.~\ref{LaNonPl}
corresponds to $\{k,l,l-r\}$ and the fourth one to $\{k,l,r\}$.

\begin{figure}[t]
  \centering
  \leavevmode
  \epsfxsize=.4\textwidth
  \epsffile[120 170 530 630]{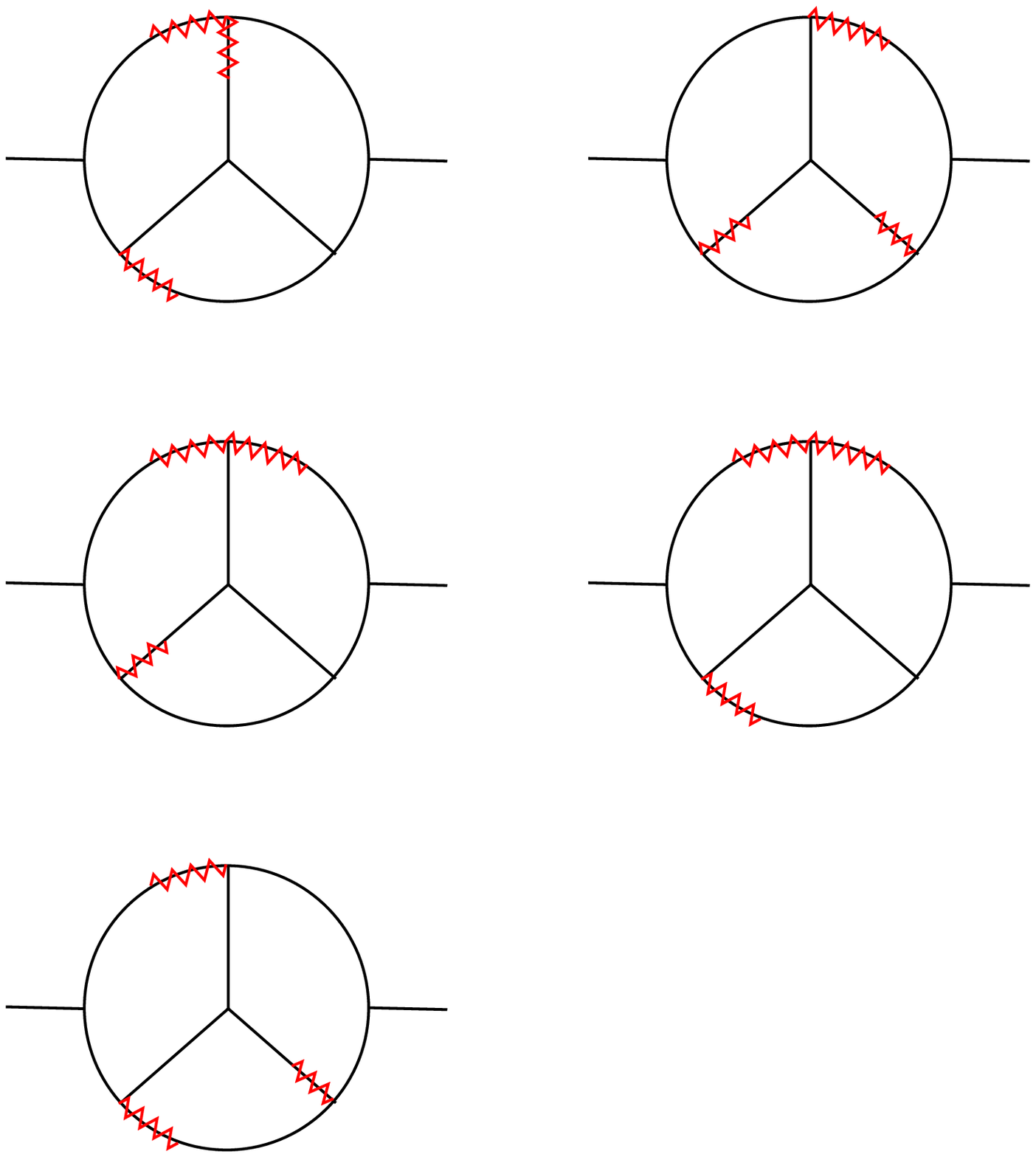}
  \caption{\label{Merc}
    Mercedes diagrams contributing to the $n_l$ part of $a_3$.}
\end{figure}

The integrals which correspond to the ``Mercedes'' graph
are shown in Fig.~\ref{Merc}. If we choose the loop momenta
$k$, $l$ and $r$ as the momenta of the three lower lines
these five diagrams appear with linear propagators of the form
$(-v\cdot k-i0)$
with momenta $\{k,k-r,l\}$, $\{r,k-l,r-l\}$, $\{k,r,k-l\}$, $\{k,r,l\}$,
$\{k,r-l,l\}$, respectively.

The next step is a reduction of all the nine types of these Feynman integrals
to master integrals by solving integration-by-parts relations \cite{Chetyrkin:1981qh}.
To do this we apply the algorithm called
{\tt FIRE} (Feynman Integral REduction)
\cite{Smirnov:2005ky,Smirnov:2006wh,Smirnov:2006vt,Smirnov:2006tz,Smirnov:2007iw,FIRE}
which is based on an extension of the classical Buchberger
algorithm to construct Gr\"obner bases (see, e.g., Ref.~\cite{Buch}).

Similarly to other  approaches, we work in a given {\em sector}, i.e. a domain
of integer indices $a_i$ where some indices are positive and the rest of the
indices are non-positive. The aim is to express any integral from the sector in
terms of master integrals of this sector {\em and} integrals from lower
sectors, where at least one more index is non-positive. It turns out that in
the higher sectors (with a small number of non-positive indices) the
corresponding $s$-basis
\cite{Smirnov:2006wh,Smirnov:2006vt,Smirnov:2006tz,Smirnov:2007iw}
(a kind of a Gr\"obner basis) can be
constructed easily (and, in most cases, even automatically).

In the opposite situation where a lot of non-positive indices occur, $s$-bases
are constructed not so easily. Usually there is the possibility to explicitly
perform an integration over some loop momentum for general value of $\ep$ with
results in terms of gamma functions. A straightforward way to do this leads to
multiple summations and turns out to be impractical. An advanced strategy
within {\tt FIRE} is to use $s$-bases for some regions of indices corresponding
to a subintegral over such a loop momentum in order to reduce these indices to
their boundary values. Then it is sufficient to use explicit integration
formulae only for the boundary values.
Integrals which are obtained from initial integrals by an explicit
integration over a loop momentum in terms of gamma function
usually involve a propagator with an analytic regularization by the shift
$\ep$ or $2\ep$.
After this integration we obtain a two-loop reduction problem with seven
indices which is then solved by {\tt FIRE}.

Finally, after using Gr\"obner bases in higher sectors and an explicit
integration in
lower sectors, it is still necessary to solve the reduction problem in a
relatively small number of intermediate sectors. In these cases we turn to
Laporta's algorithm~\cite{Laporta:1996mq,Laporta:2001dd} implemented as part
of {\tt FIRE}.

To reduce by {\tt FIRE} all the integrals contributing to the $n_l$ part of
$a_3$
it took around ten days on a 2.3 GHZ Opteron computer with 8~GB operative
memory.
The most complicated integral in this reduction is $F(1,\ldots,1,1,-4,1,0)$
corresponding the nonplanar graph in Fig.~\ref{LaNonPl} where the index of the
irreducible numerator is zero and one of the static propagators is raised to
the power $-4$. All other indices are equal to 1.


\section{Evaluating master integrals}

After using {\tt FIRE} we know the master integrals.
The number of the master integrals appearing in the $n_l$ part of $a_3$
is around one hundred. For their evaluation
we used the Mellin--Barnes (MB) technique which
is based on a replacement of a sum of terms raised
to some power by their products in some powers, at the cost of introducing
additional integrations over contours of a complex plane.
Then one takes explicitly all other integrations (over loop momenta and/or over
alpha/Feynman parameters) and is left with a multiple MB integral.

For planar diagrams, experience shows that a minimal number of
MB integrations is achieved if one introduces them loop by loop,
i.e. one derives a MB representation for a one-loop subintegral,
inserts it into a higher two-loop integral, etc.

Consider, for example, the dimensionally regularized Feynman integral
corresponding to the first graph in
Fig.~\ref{LaNonPl} with the linear propagator $(-v\cdot k+i0)^{-a_9}$.
Let us denote it by $F(a_1,\ldots,a_{11})$.
A straightforward implementation of the loop-by-loop strategy
leads to a six-fold MB representation which reads
\bea
F(a_1,\ldots,a_{11})
= \frac{\left(i\pi^{d/2} \right)^3
  (-1)^{a_9}2^{a_{10,11}-1}}{(v^2)^{a_{9,10,11}/2}}
&& \nn \\ &&  \hspace*{-68mm}
\times \frac{(-q^2)^{6-a_{1,\ldots,8}-a_{9,10,11}/2-3\ep}}
{\prod_{i=3,4,7,8,11} \Gm(a_i)\Gm(4  - a_{3,4,8,11} - 2 \ep)}
\nn \\ &&  \hspace*{-68mm}\times
\frac{1}{(2\pi i)^6} \int_{-i\infty}^{+i\infty}
\prod_{j=1}^5 \left( \Gm(-z_j) \dd z_j\right)
\frac{\Gm(-2 z_6)}{\Gm(a_{10} - 2 z_6) }
\nn \\ &&  \hspace*{-68mm} \times
\frac{\Gm(1/2 - z_1)
\Gm(1/2 - a_9/2 + z_1)}
{\Gm(1/2 + a_9/2 + z_1)\Gm(1/2 - a_9/2 - z_1)}
\nn \\ &&  \hspace*{-68mm} \times
\frac{\Gm(4 - a_{2,3,4,5,8}- a_{10,11}/2 - 2 \ep + z_1 - z_{2,3})}
{\Gm(8 -a_{1,\ldots,9}- a_{10,11}/2  - 4 \ep + z_1 - z_{2,3}) }
\nn \\ &&  \hspace*{-68mm} \times
\frac{\Gm(-6 + a_{1,\ldots,8} + a_{9,10,11}/2 +3 \ep + z_{2,3})}
{\Gm(a_{1,2,3,4,5,7,8}+ a_{10,11}/2 + 2 \ep-4 + z_{1,2,3,4})}
\nn \\ &&  \hspace*{-68mm} \times
\frac{\Gm(6 - a_{1,2,3,4,5,7,8} - a_{9,10,11}/2 - 3 \ep - z_{2,3,4}) }
{\Gm(a_6 - z_4)\Gm(a_5 - z_5)}
\nn \\ &&  \hspace*{-68mm} \times
\frac{\Gm(a_{2,3,4,5,7,8} + a_{10,11}/2 + 2 \ep-4 + z_{1,2,3,4})}
{\Gm(6-a_{2,3,4,5,7,8,10} - a_{11}/2  - 3 \ep - z_2 + z_6)}
\nn \\ &&  \hspace*{-68mm} \times
\frac{\Gm(2 - a_6 - a_9/2 - \ep + z_{1,4})\Gm(a_5 + z_{3,4} - z_5)}
{\Gm(a_{2,3,4,8} +a_{11}/2+ \ep-2 + z_{2,5,6})}
\nn \\ &&  \hspace*{-68mm} \times
\Gm(a_4 + z_{2,5}) \Gm(a_{10}/2 + z_1 - z_6)
 \Gm(a_{11}/2 + z_6)
\nn \\ &&  \hspace*{-68mm} \times
\Gm(2 - a_{4,8}- a_{11}/2  - \ep - z_{5,6})
\nn \\ &&  \hspace*{-68mm} \times
\Gm(2- a_{3,4} - a_{11}/2  - \ep - z_2 + z_6)
\nn \\ &&  \hspace*{-68mm} \times
\Gm(2- a_{5,7} - a_{10}/2  - \ep - z_{1,4} + z_{5,6})
\nn \\ &&  \hspace*{-68mm} \times
\Gm(a_{3,4,8} + a_{11}/2  + \ep -2+ z_{2,5,6})
\label{MB348}
\,,
\eea
where $a_{3,4,8,11}=a_3+a_4+a_8+a_{11}$, $z_{1,2,3}=z_1+z_2+z_3$, etc. By
definition, any integration contour over $z_i$ should go to the right (left) of
poles of gamma functions with $+z$-dependence ($-z$-dependence).

There are two strategies for resolving the singularities in $\ep$ in MB
integrals suggested in Refs.~\cite{Smirnov:1999gc,Tausk:1999vh}
(see also Chapter~4
of~\cite{Smirnov:2004ym,Smirnov:2006ry}).
The second one was formulated algorithmically
\cite{Anastasiou:2005cb,Czakon:2005rk}, and the corresponding public code {\tt
  MB.m}~\cite{Czakon:2005rk}
has become by now a standard way to evaluate MB integrals in an expansion in
$\ep$. Using {\tt MB.m} and evaluating the resulting finite MB
integrals by corollaries of Barnes lemmas we have obtained, for example, the
following result for one our master integrals:
\begin{eqnarray}
  F(1,\ldots,1,0,1)=-\frac{(i\pi^{d/2})^3}{(-q^2)^{3+3\ep} v^2}
  \left[
    -\frac{64\pi^4}{135\ep}
  \right.\nn \\ \left.\mbox{}
    -\frac{128 \pi^4}{135} - \frac{32 \pi^2\zeta(3)}{9}
    + \frac{8\zeta(5)}{3} +O(\ep)
  \right]\,.
  \label{MI348res}
\end{eqnarray}
In contrast to a similar Feynman integral with the linear propagator
$1/(-v\cdot k-i0)$ (see Ref.~\cite{Smirnov:2008tz}) one needs, at a first step
of the resolution
of the singularities in $\ep$, an auxiliary analytic
regularization. In fact, the result in Eq.~(\ref{MI348res}) and
the corresponding one for the case of $1/(-v\cdot k-i0)^{a_9}$
(see Ref.~\cite{Smirnov:2008tz}) differ by a term which is nothing but a
residue taken when performing the initial analytical continuation to the
auxiliary parameter of analytic regularization.

The above integral is an example where products of the type
$1/(x+i0)^{a}1/(x-i0)^{b}$ appear, with $x=-v\cdot k, -v\cdot l$, etc.
If $a$ and $b$ are positive integers, these products are ill-defined. For
example, they do appear in diagrams resulting from iterations of the Coulomb
potential. However, for integrals in Eq.~(\ref{MB348}), and many other
integrals that we
encounter, in these products  at least one of the exponents is
regularized by an amount proportional to $\ep$ and/or $z$ (i.e. a
variable of the MB integration). As a consequence they are well-defined,
in the sense of analytic continuation. More explicitly, we have
\bea
\lefteqn{(x+i0)^{a}(x-i0)^{b} =}
\nonumber\\&&\mbox{}
\frac{\sin (\pi a) e^{-i\pi b}}{\sin (\pi (a+b))} (x+i0)^{a+b}
\nonumber\\&&\mbox{}
+\frac{\sin (\pi b) e^{i\pi a}}{\sin (\pi (a+b))} (x-i0)^{a+b}\,,
\eea
for general complex numbers $a$ and $b$.

All the master integrals that we encounter are expressed in terms of
$\zeta(i)$, i=2,3,4,5,6, powers of $\ln\,2$, Li$_j(1/2), j=4,5,6$, and the
constant $s_6$ (see, e.g., \cite{Maitre:2005uu}) up to transcendentality level
six.
With the assumption about the presence of these numbers in the results, one can
apply the PSLQ algorithm~\cite{PSLQ} in situations where explicit
integration over MB parameters by corollaries of Barnes lemmas is no longer
possible. If this is just a onefold MB integral then it is possible to obtain a
numerical result with more than 300 digits which are sufficient for the
PSLQ algorithm.
For twofold MB integrals, one can hope to obtain an accuracy of 50 digits
depending on the complexity of the integrand. This accuracy can also be
sufficient for a successful application of the PSLQ algorithm, at least when a
given integral possesses homogeneous transcendentality.
Unfortunately, sufficient criteria about homogeneous transcendentality are not
known at the moment so that this property can be seen only experimentally,
by considering lower terms of the expansion in $\ep$.


\section{Preliminary results and perspectives}

The three-loop correction to the static quark potential can
conveniently be parameterized in the form
\begin{eqnarray}
  a_3 &=& a_3^{(3)} n_l^3 + a_3^{(2)} n_l^2 + a_3^{(1)} n_l + a_3^{(0)}\,,
\end{eqnarray}
where $n_l$ denotes the number of massless quarks. Using the techniques
described above we evaluated the coefficients $a_3^{(3)}$ and $a_3^{(2)}$
which read
\begin{eqnarray}
  a_3^{(3)} &=& - \left(\frac{20}{9}\right)^3 T_F^3
  \,,\nonumber\\
  a_2^{(2)} &=&
  \left(\frac{12541}{243}
    + \frac{368\zeta(3)}{3}
    + \frac{64\pi^4}{135}
  \right) C_A T_F^2
  \nonumber\\&&\mbox{}
  +
  \left(\frac{14002}{81}
    - \frac{416\zeta(3)}{3}
  \right) C_F T_F^2
  \,.
\end{eqnarray}
The $n_l^3$ contribution together with the $C_A T_F^2$ part has already been
presented in Ref.~\cite{Smirnov:2008tz}.

The coefficient $a_3^{(1)}$ is also reachable. Only four constants
contributing to the higher-order expansion in $\ep$ of some master integrals
are not known analytically at the moment.



\vspace*{2mm}


\noindent
{\bf Acknowledgements.}
We thank A. Penin for useful communication.
This work was supported by RFBR, grant 08-02-01451, and by DFG through project
SFB/TR~9.


\end{document}